\documentclass[conference]{IEEEtran}
\IEEEoverridecommandlockouts

\usepackage{cite}
\usepackage{amsmath,amssymb,amsfonts}
\usepackage{algorithmic}
\usepackage{algorithm}
\usepackage{graphicx}
\usepackage{textcomp}
\usepackage{xcolor}
\usepackage{booktabs}
\usepackage{array}
\usepackage{multirow}
\usepackage{url}
\usepackage{balance}
\usepackage[hidelinks]{hyperref}

\graphicspath{{figs/}}

\def\BibTeX{{\rm B\kern-.05em{\sc i\kern-.025em b}\kern-.08em
    T\kern-.1667em\lower.7ex\hbox{E}\kern-.125emX}}
    
\newcommand{\name}{\textit{FedGuard-DC }}
\newcommand{\namenospace}{\textit{FedGuard-DC}}

\begin{document}

\title{\namenospace: Privacy-Preserving Federated Load Forecasting and Cyber-Attack Detection for Data-Center Loads in Transmission Systems}

 \author{
 \IEEEauthorblockN{Md Kibria Saroare, Md Rubel Ahmed}
  Louisiana Tech University
 \IEEEauthorblockA{\{msa078,mahmed\}@latech.edu}
 }

\maketitle

\begin{abstract}
The rapid growth of large data-center (DC) loads is creating new challenges for power-system visibility, privacy, and cyber-physical security. System operators need accurate short-term information about these fast-varying loads, while DC operators may avoid sharing raw megawatt measurements because they can reveal sensitive workload and utilization patterns. This paper presents \namenospace, a federated learning (FL) framework for privacy-preserving DC load forecasting and local false-data-injection attack (FDIA) detection. Each DC trains a dual-head model on its own measurements, where a shared encoder supports both a forecasting head and a reconstruction head. A calibrated anomaly score combines forecast residual and reconstruction error to detect corrupted measurements locally. Raw measurements and absolute MW demand remain at each DC, while only model updates, normalized forecast outputs, and scalar attack flags are shared with the global controller. Optional differential privacy and robust trimmed-mean aggregation are included to evaluate privacy--utility behavior and poisoned-client resilience. The framework is validated using phasor-domain simulation data from four large DC loads rated between 150 and 350~MW integrated into the IEEE 39-bus New England system. Results show a 0.5~s-ahead normalized forecast RMSE of 0.023--0.038~pu, compared with 0.32--0.34~pu for persistence. \name detects FDIA with ROC-AUC of 0.979, $F_{1}=0.930$, and precision of 0.988, while robust aggregation reduces the poisoned-client RMSE impact from 0.042 to 0.035~pu.
\end{abstract}

\begin{IEEEkeywords}
Data center, federated learning, FDIA, load forecasting, cyber-security, IEEE 39-bus.
\end{IEEEkeywords}

\vspace{-10pt}
\section{Introduction}
\label{sec:intro}
Large data centers are becoming an important operating concern for modern power systems. Driven by cloud computing, artificial-intelligence (AI) services, and high-performance digital infrastructure, their electricity use is growing rapidly. Global data-center electricity consumption was about 415~TWh in 2024 and is projected to reach around 945~TWh by 2030~\cite{iea2025energyai}. In the United States, data-center electricity use was estimated at 176~TWh in 2023 and may rise to 325--580~TWh by 2028~\cite{lbnl2024usdatacenter}. Recent reliability studies also identify emerging large loads, including data centers, as a planning and operational challenge for the bulk grid~\cite{nerc2025ltra}.

A data center can appear to the grid as a concentrated load of hundreds of megawatts. Its demand may change quickly because of server utilization, AI workload variation, cooling operation, UPS interfaces, and backup resources. System operators therefore need accurate short-term forecasts and efficient monitoring capabilities for dispatch, congestion management, voltage-security assessment, and situational awareness~\cite{saroare2026gridstream}. However, raw megawatt measurements may reveal utilization levels, tenant activity, and workload patterns, making data-center operators reluctant to share them directly.

Cyber-physical security further complicates this problem. Data centers depend on sensors, local controllers, supervisory systems, and communication links, which can expose measurement streams to data-integrity attacks. False-data-injection attacks can corrupt measurements used by forecasting and monitoring tools while remaining difficult to detect using conventional bad-data logic~\cite{liu2011fdi}. In a data-center-rich grid, compromising one large-load measurement stream can distort the operator's view of demand and affect operator actions.

Federated learning is suitable for this setting because it enables shared model training without transferring raw local data~\cite{mcmahan2017fedavg}. Differential privacy can further reduce information leakage from exchanged model updates~\cite{dwork2014dp}. Existing FL-based studies address power-system FDIA detection and short-term load forecasting in several settings~\cite{li2022secure,tran2023crosssilo,lin2022incentive,lin2024privacy,kesici2024vertical,fernandez2022privacy,fekri2022rnn,husnoo2023secure}, but most treat forecasting and attack detection separately and focus on residential, feeder-level, distribution-network, or general grid measurements. The combined problem of privacy-preserving forecasting and local FDIA detection for large, fast-varying data-center loads remains insufficiently studied.

This paper proposes \namenospace, a federated framework for privacy-preserving data-center load forecasting and FDIA detection. In the proposed framework, each data center trains a dual-head local model in which a shared encoder supports both a forecasting head and a reconstruction head. FDIA is detected locally by a calibrated anomaly score that combines forecast residual and reconstruction error. This keeps raw measurements and absolute MW demand private, while only updated model weights are shared with the global controller.
The main contributions are:
\begin{enumerate}
\item A dual-head local model for joint short-term data-center load forecasting and measurement reconstruction.
\item A calibrated FDIA detection score based on forecast residual and reconstruction error, with automatic threshold selection.
\item Validation using phasor-domain simulation data from four large data-center loads integrated into the IEEE 39-bus New England system, including forecasting, FDIA detection, privacy--utility, and poisoned-client robustness studies.
\end{enumerate}

\noindent
The remainder of this paper is organized as follows. Section~\ref{sec:related} reviews related work. Section~\ref{sec:framework} presents the proposed framework. Section~\ref{sec:case} describes the case-study setup. Section~\ref{sec:results} discusses the results. Section~\ref{sec:conclusion} concludes the paper.

\vspace{-3pt}
\section{Related Work}
\label{sec:related}

Recent FL-based cyber-security studies show that distributed learning can support power-system attack detection while limiting direct data sharing. Secure aggregation, cross-silo FL, incentive-based edge learning, privacy-preserving model exchange, and vertical FL have been applied to FDIA detection in transmission, smart-grid, and distribution-network settings~\cite{li2022secure,tran2023crosssilo,lin2022incentive,lin2024privacy,kesici2024vertical}. Related work also considers federated anomaly detection, poisoned-update defense, and distributed FDIA localization~\cite{jithish2023distributed,li2024poisoning,kececi2025localization}. These studies support FL for privacy-aware cyber-physical monitoring but mainly address grid, smart-meter, distribution, and renewable-energy data.

FL has also been used for short-term load forecasting in residential, smart-meter, building-level, and heating-demand applications. Recent work addresses privacy-preserving residential forecasting, recurrent-network-based distributed forecasting, clustered residential forecasting, personalized consumer forecasting, heterogeneous-load forecasting and secure residential forecasting~\cite{fernandez2022privacy,fekri2022rnn,briggs2022residential,wang2023personalized,qu2023heterogeneous,husnoo2023secure}. These studies usually focus on smaller or slower-varying demand profiles and treat forecasting and attack detection as separate tasks. In contrast, this paper combines short-term forecasting, reconstruction-based integrity monitoring, local FDIA detection, privacy-preserving FL, and robust aggregation in one framework for data-center-rich power systems.

\vspace{-3pt}
\section{\name Framework}
\label{sec:framework}

\begin{figure}[!t]
   \centering
   \includegraphics[height=0.38
   \textheight,keepaspectratio]{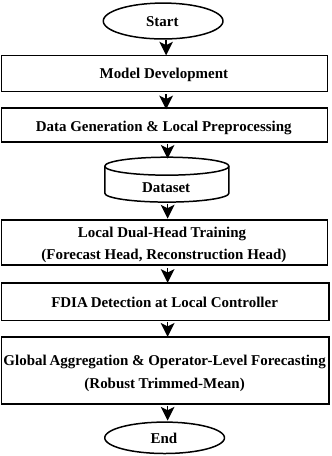}
   \caption{Overall workflow of the proposed \name framework. The framework combines privacy-preserving load forecasting and FDIA detection within a single federated model, where a shared encoder drives both a forecast head and a reconstruction head locally at each data center.}
   \vspace{-5pt}
    \label{fig:workflow}
\end{figure}

The overall \name workflow is shown in Fig.~\ref{fig:workflow}. Each data center operates as a local FL client that preprocesses its own voltage and power measurements, trains the dual-head forecasting-reconstruction model, and computes the FDIA anomaly score locally. Model updates are sent to the global controller for federated aggregation, while normalized forecast outputs and scalar attack flags provide operator-level monitoring without exposing raw measurements, private scaling factors, or absolute MW demand. Major components of the framework are described as follows.

\medskip
\subsubsection{Data-Center Power Architecture}
\begin{figure}[!t]
    \centering
    \includegraphics[width=0.95\columnwidth]{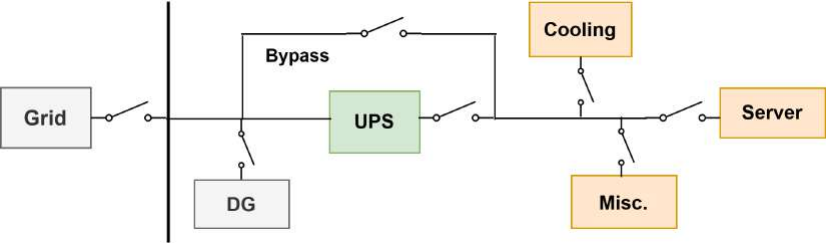}
    \caption{Typical data-center power architecture. Grid power feeds an internal UPS bus via a main breaker (with bypass path); IT servers, cooling, and miscellaneous loads share the bus, with a diesel generator (DG) providing standby backup.}
    \vspace{-15pt}
    \label{fig:dc_arch}
\end{figure}

The electrical architecture of the data-center used in this study follows the DC-CLM representation developed in our previous work~\cite{saroare2026dcclm} and is shown in Fig.~\ref{fig:dc_arch}. A large data center is represented by IT/server, cooling, and various auxiliary loads, together with a UPS interface and backup generator. The IT branch includes servers, storage, networking equipment and AI accelerators; the cooling branch includes chillers, pumps, fans, and other thermal-management equipment; and the miscellaneous branch represents facility-level auxiliary demand. Under normal operation, the grid supplies the aggregate demand for the data-center through the UPS path. During voltage or frequency disturbances, the UPS can isolate protected internal loads from the external grid and maintain service until acceptable grid conditions return, while the backup generator supports the facility during extended outages. This structure captures the main sources of data-center load dynamics considered in this paper: fast IT workload variation, cooling-system demand, UPS switching, and backup-resource support.

\medskip
\subsubsection{Threat Model}
Two threat types are considered. The first is a measurement-level FDIA adversary that corrupts local input windows through scaling, ramp/drift, replay, spike/noise, or stealthy bias attacks. The second is a poisoned-client adversary that compromises one local controller and sends a malicious model update to the aggregator. The proposed framework addresses these threats through local anomaly detection, privacy-preserving model sharing, optional differential privacy, and robust trimmed-mean aggregation.

\medskip
\subsubsection{Measurement Window and Local Normalization}
Each DC~$k$ exposes a time series of three-phase terminal voltages $\mathbf{v}_k(t)=[v_a,v_b,v_c]$ (pu) and total active power $P_k(t)$ (W), the latter comprising IT, cooling, and overhead components. The local controller forms a sliding measurement window
\begin{equation}
    \mathbf{x}_k(t)=\bigl[v_a(\tau),v_b(\tau),v_c(\tau),p_k(\tau)\bigr]_{\tau=t-W+1}^{t}\in\mathbb{R}^{W\times 4},
    \label{eq:window}
\end{equation}
where $p_k=P_k/s_k$ is power normalized by the locally held scale $s_k=\max_{t}P_k(t)$. The scale $s_k$ \emph{never leaves} DC~$k$, which is the cornerstone of the privacy design: recovering any DC's absolute MW demand from the shared model would require its private scale.

\medskip
\subsubsection{Dual-Head Local Model}
All DCs train an identical architecture on their private data. The flattened window $\mathbf{x}\in\mathbb{R}^{4W}$ passes through a shared encoder:
\begin{align}
    \mathbf{h}&=\mathrm{ReLU}(\mathbf{W}_{1}\mathbf{x}+\mathbf{b}_{1}),\label{eq:enc1}\\
    \mathbf{z}&=\mathrm{ReLU}(\mathbf{W}_{2}\mathbf{h}+\mathbf{b}_{2}),\label{eq:enc2}
\end{align}
producing a latent code $\mathbf{z}$, on which two independent heads operate as:
\begin{equation}
    \hat{y}=\mathbf{w}_{f}^{\top}\mathbf{z}+b_{f},\qquad
    \hat{\mathbf{x}}=\mathbf{W}_{d}\mathbf{z}+\mathbf{b}_{d}.
    \label{eq:heads}
\end{equation}
The forecast head $\hat{y}$ predicts normalized power $H$ steps ahead. The reconstruction head $\hat{\mathbf{x}}$ rebuilds the input window, functioning as an autoencoder that memorizes the statistics of clean measurements. The two objectives are jointly minimized:
\begin{equation}
    \mathcal{L}=\frac{1}{B}\sum_{i=1}^{B}(\hat{y}_{i}-y_{i})^{2}+\frac{\lambda}{B}\sum_{i=1}^{B}\|\hat{\mathbf{x}}_{i}-\mathbf{x}_{i}\|_{2}^{2}.
    \label{eq:loss}
\end{equation}
Joint training forces the encoder to satisfy two tasks: (i) predict the future load, and (ii) reproduce the clean input manifold simultaneously. An attack distorts the input, disrupting both tasks and making the detector powerful.

\medskip
\subsubsection{Calibrated FDIA Detection Score}
At inference, DC~$k$ computes two error terms:
\begin{equation}
    e_{\mathrm{rec}}=\frac{1}{4W}\|\hat{\mathbf{x}}-\mathbf{x}\|_{2}^{2},\qquad
    e_{\mathrm{fc}}=(\hat{y}-y)^{2}.
    \label{eq:errors}
\end{equation}
Because these two error terms live on different scales, both are standardized against clean-data statistics $(\mu_{r},\sigma_{r})$ and $(\mu_{f},\sigma_{f})$ estimated on an attack-free calibration set. The composite anomaly score is
\begin{equation}
    s=\alpha\,\frac{e_{\mathrm{rec}}-\mu_{r}}{\sigma_{r}}+\beta\,\frac{e_{\mathrm{fc}}-\mu_{f}}{\sigma_{f}}.
    \label{eq:score}
\end{equation}
A window is flagged as attacked when $s>\tau^{\star}$. Rather than setting $\tau^{\star}$ by hand, both the score weights $(\alpha,\beta)$ and the threshold are selected automatically by exhaustive grid search on a labeled validation mixture to maximize $F_{1}$:
\begin{equation}
    (\alpha^{\star},\beta^{\star},\tau^{\star})=\arg\max_{\alpha,\beta,\tau}F_{1}\!\bigl(\mathbf{1}[s(\alpha,\beta)>\tau]\bigr).
    \label{eq:gridsearch}
\end{equation}


\medskip
\subsubsection{Federated Training, Privacy, and Robustness}

\begin{algorithm}[!t]
\caption{Training and Real-Time FDIA Monitoring}
\label{alg:fedguard}
\begin{algorithmic}[1]
\REQUIRE Clients $\{k\}$, rounds $R$, epochs $E$, clip $C$, noise $\sigma$
\STATE Initialize global parameters $\boldsymbol{\theta}$
\FOR{$r=1$ to $R$}
    \FOR{each DC $k$ in parallel}
        \STATE $\boldsymbol{\theta}_{k}\leftarrow\textsc{LocalTrain}(\boldsymbol{\theta},\mathrm{data}_{k},E)$ \hfill \COMMENT{Eq.~\eqref{eq:loss}}
        \STATE $\Delta_{k}\leftarrow\boldsymbol{\theta}_{k}-\boldsymbol{\theta};\ \tilde{\Delta}_{k}\leftarrow\textsc{DP}(\Delta_{k},C,\sigma)$ \hfill \COMMENT{Eq.~\eqref{eq:dp}}
    \ENDFOR
    \IF{robust mode}
        \STATE $\boldsymbol{\theta}\leftarrow\boldsymbol{\theta}+\textsc{TrimmedMean}(\{\tilde{\Delta}_{k}\})$
    \ELSE
        \STATE $\boldsymbol{\theta}\leftarrow\boldsymbol{\theta}+\sum_{k}w_{k}\tilde{\Delta}_{k},\ w_{k}\propto n_{k}\sqrt{\ell_{k}}$
    \ENDIF
\ENDFOR
\STATE Calibrate $(\mu_{r},\sigma_{r},\mu_{f},\sigma_{f})$ on clean data
\STATE Auto-tune $(\alpha^{\star},\beta^{\star},\tau^{\star})$ \hfill \COMMENT{Eq.~\eqref{eq:gridsearch}}
\STATE Monitor: flag window if $s(\alpha^{\star},\beta^{\star})>\tau^{\star}$
\end{algorithmic}
\end{algorithm}
\vspace{-10pt}

The full training and real-time monitoring protocol is given in Algorithm~\ref{alg:fedguard}, which proceeds in three phases: federated model training, post-training calibration, and real-time detection.

In the training phase (lines~1--12), the global controller initializes shared parameters $\boldsymbol{\theta}$ and runs $R$ communication rounds. In each round, every DC~$k$ independently performs $E$ epochs of local training using the joint loss in Eq.~\eqref{eq:loss}, producing updated local parameters $\boldsymbol{\theta}_k$. The local update vector $\Delta_{k}=\boldsymbol{\theta}_{k}-\boldsymbol{\theta}$ captures the change induced by that DC's private data. Before transmission, each client update is clipped and perturbed with Gaussian noise:
\begin{equation}
    \tilde{\Delta}_{k}=\Delta_{k}\cdot\min\!\Bigl(1,\frac{C}{\|\Delta_{k}\|_{2}}\Bigr)+\mathcal{N}(\mathbf{0},\sigma^{2}C^{2}\mathbf{I}),
    \label{eq:dp}
\end{equation}
with clipping bound $C$ and noise multiplier $\sigma$. The global controller aggregates using a variability-weighted rule: each client's contribution is proportional to both its data volume $n_{k}$ and the square root of its local training loss $\ell_{k}$, so harder-to-model (more volatile) DCs exert more influence on the global model. Under a poisoning threat, the operator switches to coordinate-wise trimmed-mean aggregation with $t=1$, which discards the largest and smallest client updates at each coordinate and averages the remaining two updates, thereby reducing the influence of a single amplified malicious update.

In the calibration phase (lines~13--14), after training converges, clean-data statistics $(\mu_r, \sigma_r, \mu_f, \sigma_f)$ are estimated on an attack-free validation set to standardize the two error terms in Eq.~\eqref{eq:score}. The score weights $(\alpha, \beta)$ and decision threshold $\tau^\star$ are then jointly optimized via grid search to maximize $F_1$ as in Eq.~\eqref{eq:gridsearch}, removing the need for manual threshold tuning.

In the real-time monitoring phase (line~15), each local controller continuously evaluates the composite anomaly score $s(\alpha^\star, \beta^\star)$ on incoming measurement windows and raises a binary attack flag whenever $s > \tau^\star$. The operator receives only model updates, normalized forecast outputs, and scalar attack flags, while raw measurements, private scaling factors, and absolute MW demand remain local.


\section{Case Study Setup}
\label{sec:case}

\begin{figure}[!t]
    \centering
    \includegraphics[width=\columnwidth]{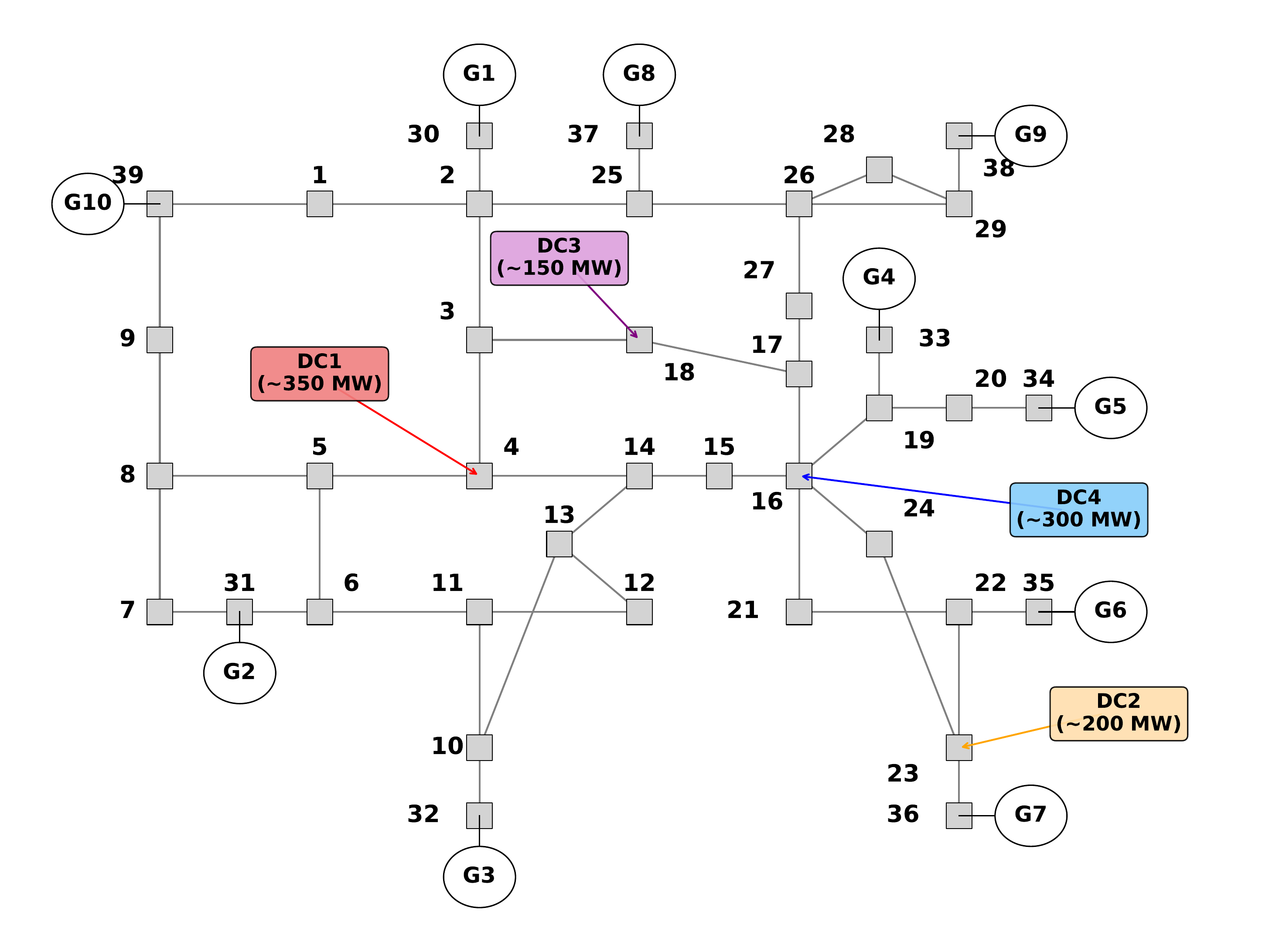}
    \vspace{-15pt}
    \caption{Modified IEEE 39-bus New England system with four embedded data-center loads.}
    \label{fig:ieee39}
    \vspace{-15pt}
\end{figure}

The proposed framework is evaluated on the IEEE 39-bus New England system~\cite{athay1979ieee39}, a standard 10-machine, 345~kV transmission benchmark with 34 transmission lines, 12 transformers, and 19 load buses. Four of the existing constant loads are replaced by the DC models described in Table~\ref{tab:dcloads}. The DCs are connected at bus~4 (DC1, $\approx$~350~MW), bus~23 (DC2, $\approx$~200~MW), bus~18 (DC3, $\approx$~150~MW), and bus~16 (DC4, $\approx$~300~MW), as shown in Fig.~\ref{fig:ieee39}. These placements stress different zones of the network and produce diverse voltage profiles at the measurement terminals.

\begin{table}[b]
\centering
\vspace{-18pt}
\caption{Data-Center Loads Embedded in the IEEE 39-Bus System}
\label{tab:dcloads}
\renewcommand{\arraystretch}{1.15}
\setlength{\tabcolsep}{4pt}
\begin{tabular}{lcccccc}
\toprule
DC & Bus & Rating (MW) & Peak (MW) & CoV & IT \% & Cool \% \\
\midrule
DC1 & 4   & 350 & 320.6 & 0.42 & 64 & 30 \\
DC2 & 23 & 200 & 180.7 & 0.46 & 65 & 30 \\
DC3 & 18     & 150 & 136.6 & 0.47 & 65 & 30 \\
DC4 & 16 & 300 & 274.1 & 0.43 & 64 & 30 \\
\bottomrule
\end{tabular}
\end{table}

\medskip
\paragraph{Data Generation and Preprocessing}
Each DC provides 120,474 samples logged at 1~kHz over 120~s from a phasor-domain simulation of the IEEE 39-bus system in MATLAB/Simulink. The initial 5~s simulation transient is discarded, and the series is downsampled by a factor of ten to a 10~ms resolution, yielding approximately 11,473 sliding windows per DC. Input windows use $W=40$ steps, and the forecast horizon is $H=50$ steps, corresponding to a 0.5~s lead time. A 70/30 chronological train/test split is applied per DC (8,031 training and 3,442 test windows), and no raw data is shared between DCs at any stage. The resulting per-DC normalized load profiles in Fig.~4 exhibit pronounced high-frequency duty-cycling across the four data centers.

\begin{figure}[!t]
    \centering
    \includegraphics[width=\columnwidth]{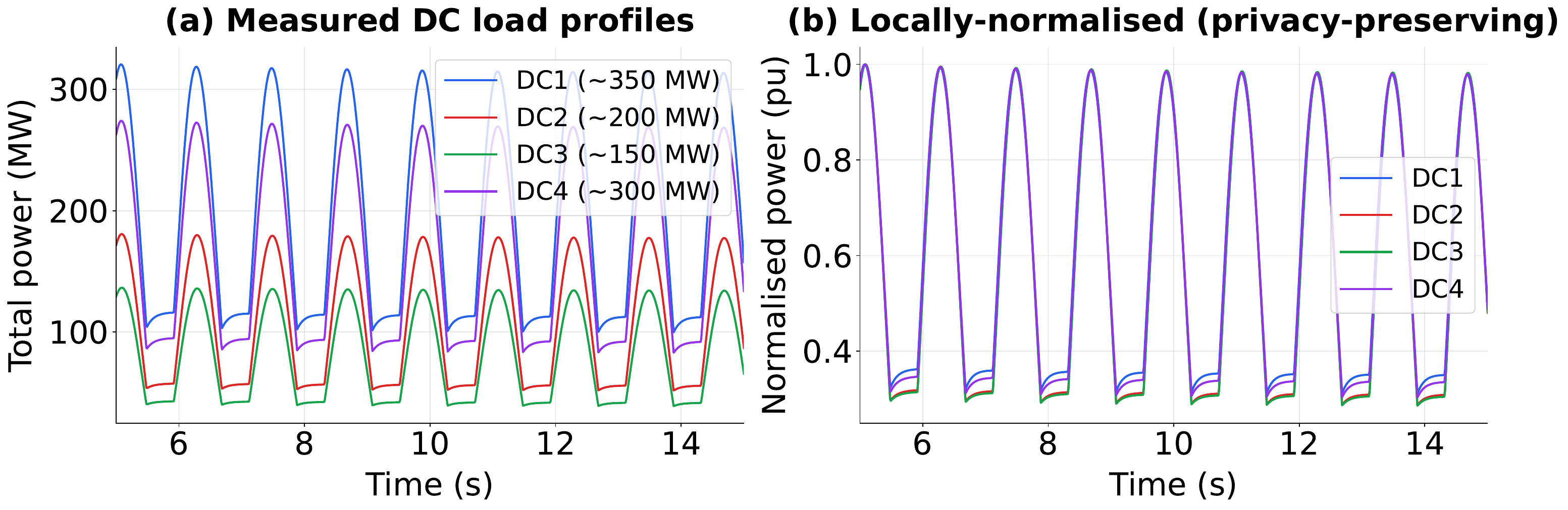}
    \caption{Data-center load profiles obtained from the phasor-domain simulation: (a) simulated active-power demand and (b) locally normalized profiles used by FedGuard-DC.}
    \label{fig:loadprof}
    \vspace{-15pt}
\end{figure}

\medskip
\paragraph{Hyperparameters, Baselines, and Metrics}
The shared encoder has hidden width 256 and latent dimension 128; $\lambda=0.5$, learning rate $2\times10^{-3}$ (Adam), batch size 256, $R=150$ communication rounds, and $E=2$ local epochs per round. For detection, $(\alpha,\beta)$ are jointly tuned over a $20\times20$ grid with $(\alpha,\beta)\in[0.1,5]$, yielding $\alpha^{\star}=3.97,\beta^{\star}=0.79$. The DP clipping norm is $C=5.0$; the noise multiplier is $\sigma=0$ unless explicitly stated otherwise. The persistence baseline predicts the last observed value in the window. Forecasting is measured via RMSE, MAE, and MAPE in normalized pu values. Detection is measured via ROC-AUC, $F_{1}$, accuracy, precision, recall, and per-attack-type recall on a test mixture with $50\%$ attacked windows.


\section{\name Evaluation and Discussion}
\label{sec:results}

\begin{figure*}[!t]
    \centering
    \includegraphics[width=0.95\textwidth]{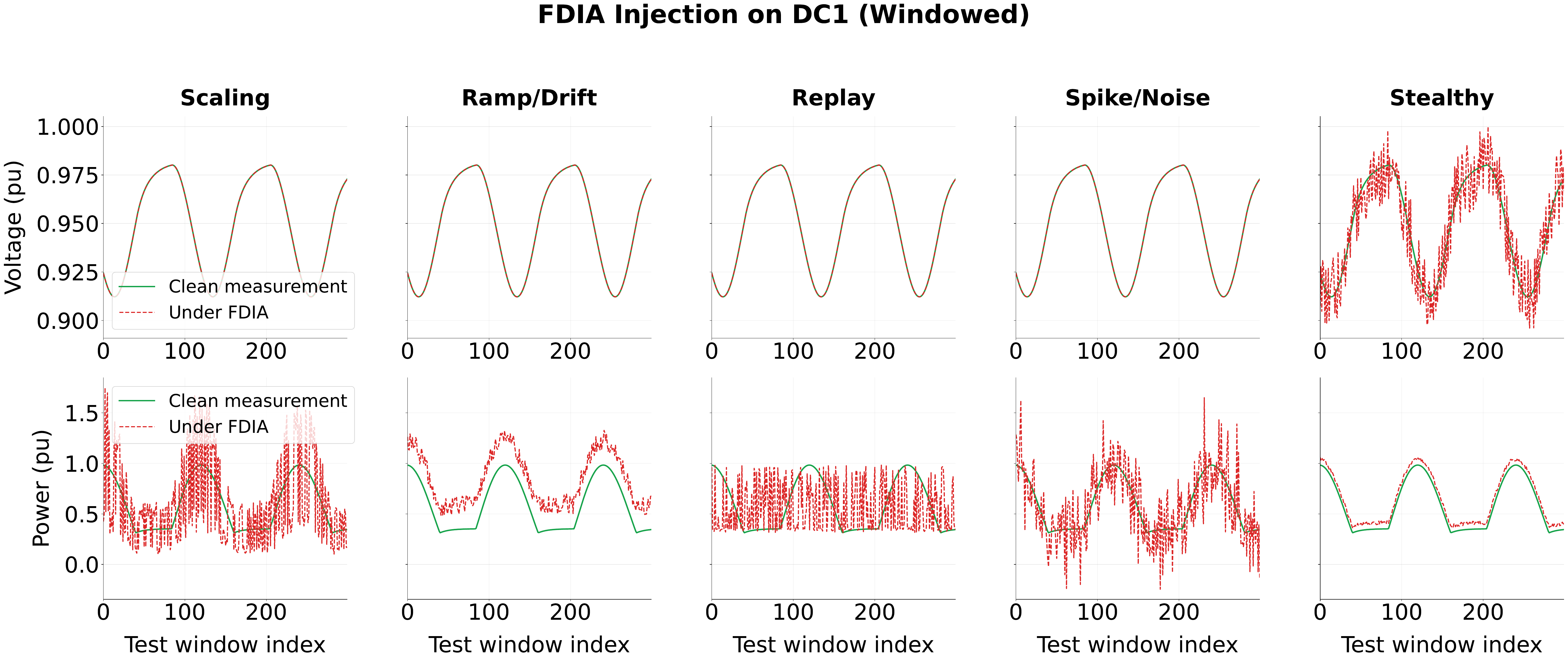}
    \caption{Five representative FDIA scenarios over a DC1 test segment (top row: voltage; bottom row: power). From left to right: scaling, ramp/drift, replay, spike/noise, and stealthy. The stealthy attack closely mimics the clean trace and is the hardest case for any anomaly detector.}
    \label{fig:attacks}
    \vspace{-15pt}
\end{figure*}

To evaluate measurement-integrity detection, five FDIA types are injected into the local data-center measurement windows, as illustrated in Fig.~\ref{fig:attacks}. The attacks include scaling, ramp/drift, replay, spike/noise, and stealthy bias. These cases represent both obvious corruptions, such as large power scaling and noise bursts, and more difficult attacks, such as replayed measurements and small coordinated bias. The stealthy case is the most challenging because its deviation is close to normal load variation. This attack set is used to test whether the proposed anomaly score can detect corrupted measurements under different levels of visibility and severity.

\medskip
\subsubsection{Load Forecasting}
Table~\ref{tab:forecast} summarizes the 0.5~s-ahead forecasting results for the four data centers. \name achieves normalized RMSE values of $0.023$--$0.038$~pu and MAPE values of $3.17$--$6.27 \%$, substantially improving over the persistence baseline of $0.32$--$0.34$~pu. This confirms that the proposed model can learn the fast duty-cycling behavior of data-center loads, where a last-observation forecast quickly becomes inaccurate. As shown in Fig.~6, the predicted traces closely follow the normalized power profiles for all four data centers. The aggregate result in Fig.~\ref{fig:aggforecast} further shows that the operator can obtain useful system-level insight without accessing individual raw MW measurements or  scaling factors.

\begin{table}[!t]
\centering
\caption{Forecasting Performance per DC (0.5~s Lead Time)}
\label{tab:forecast}
\renewcommand{\arraystretch}{1.15}
\setlength{\tabcolsep}{5pt}
\begin{tabular}{lcccc}
\toprule
DC & RMSE (pu) & MAE (pu) & MAPE (\%) & Persist. RMSE (pu)\\
\midrule
DC1 & 0.038 & 0.034 & 6.27 & 0.315\\
DC2 & 0.029 & 0.025 & 4.70 & 0.334\\
DC3 & 0.027 & 0.023 & 4.48 & 0.335\\
DC4 & 0.023 & 0.018 & 3.17 & 0.322\\
\midrule
\textbf{Avg.} & \textbf{0.029} & \textbf{0.025} & \textbf{4.66} & \textbf{0.327}\\
\bottomrule
\end{tabular}
\vspace{-5pt}
\end{table}

\begin{figure}[!t]
    \centering
    \includegraphics[width=\columnwidth]{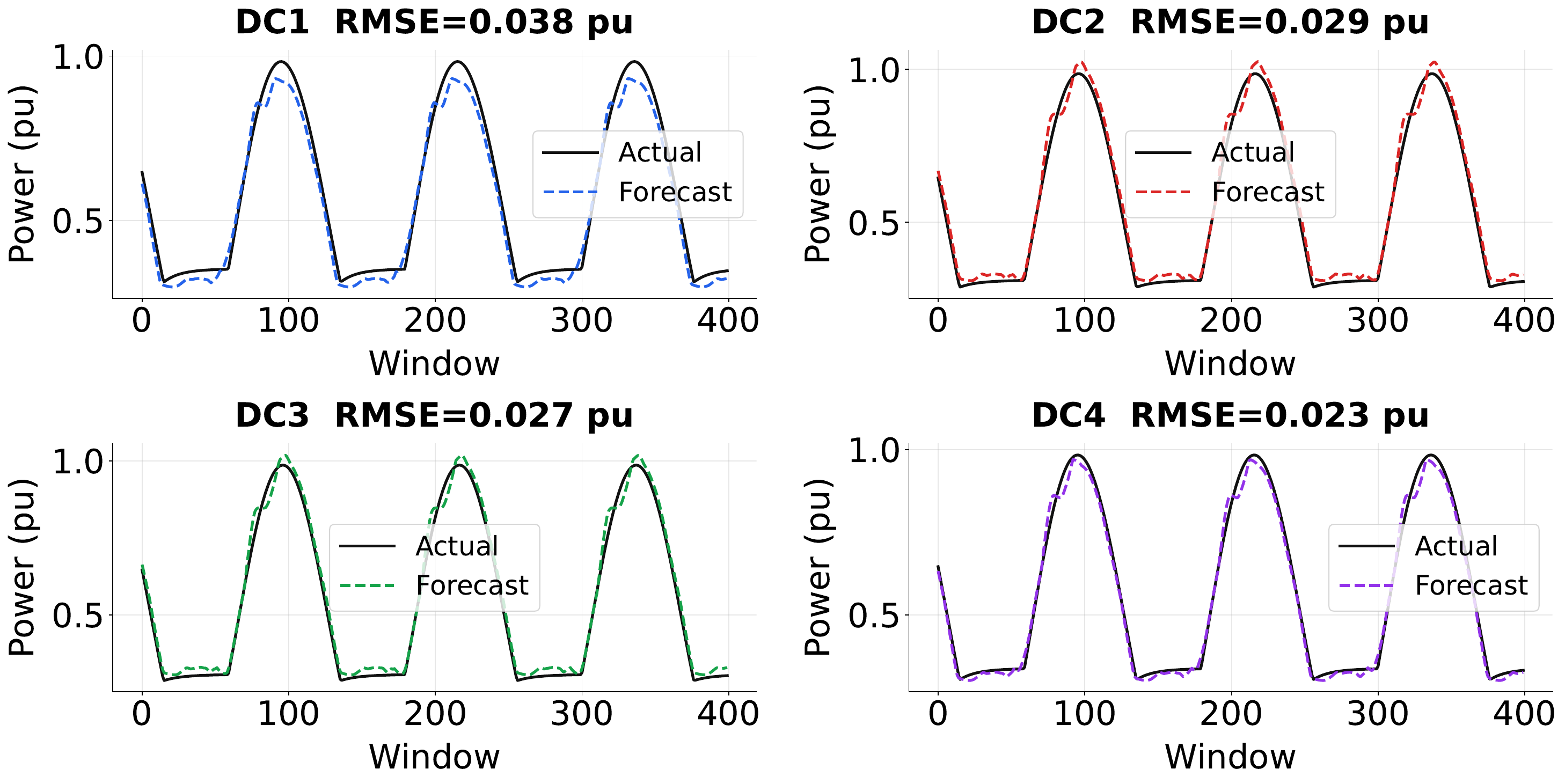}
    \caption{Actual vs.\ forecast normalized power for all four DCs over a 400-window test segment.}
    \vspace{-10pt}
    \label{fig:forecast}
    \vspace{-10pt}
\end{figure}

\begin{figure}[!t]
    \centering
    \includegraphics[width=\columnwidth]{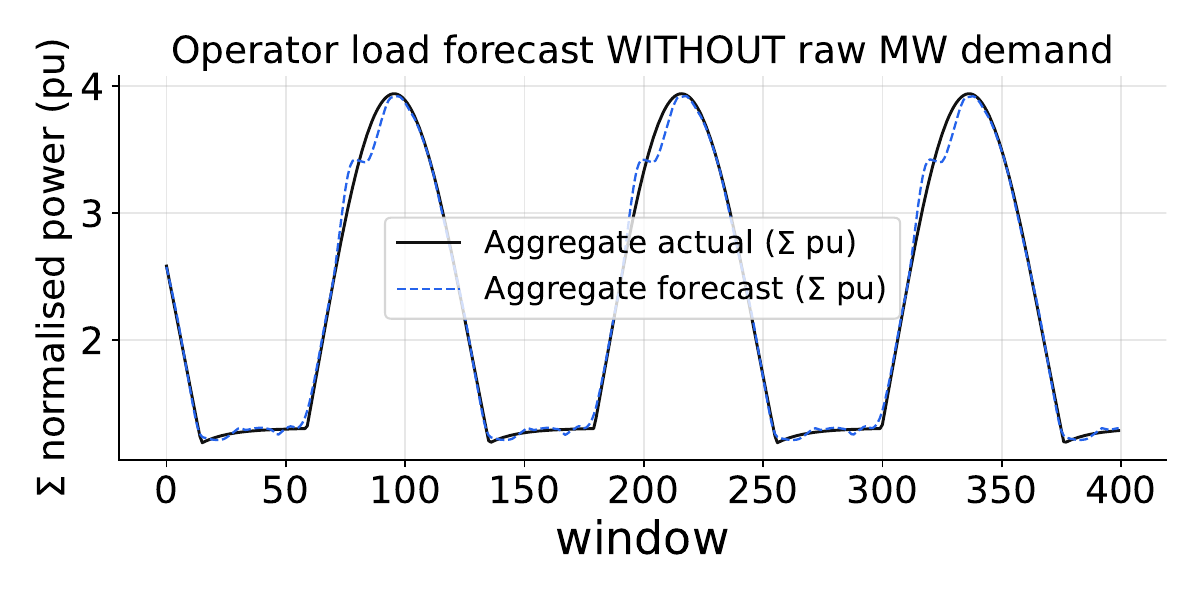}
    \vspace{-10pt}
    \caption{Operator-level aggregate load forecast ($\Sigma$~pu) without raw MW-demand disclosure.}
    \vspace{-10pt}
    \label{fig:aggforecast}
    \vspace{-10pt}
\end{figure}

\medskip
\subsubsection{Attack Detection}
With auto-tuned parameters $(\alpha^{\star}=3.97,\,\beta^{\star}=0.79,\,\tau^{\star}=13.22)$, \name achieves ROC-AUC~$=0.979$, $F_{1}=0.930$, accuracy~$=0.934$, precision~$=0.988$, and recall~$=0.879$ on the mixed test set. The high precision of 0.988 indicates that nearly all windows flagged as attacks are true attacks in the evaluated test mixture, which is desirable for an operational alarm system.

\begin{figure}[!t]
    \centering
    \includegraphics[width=\columnwidth]{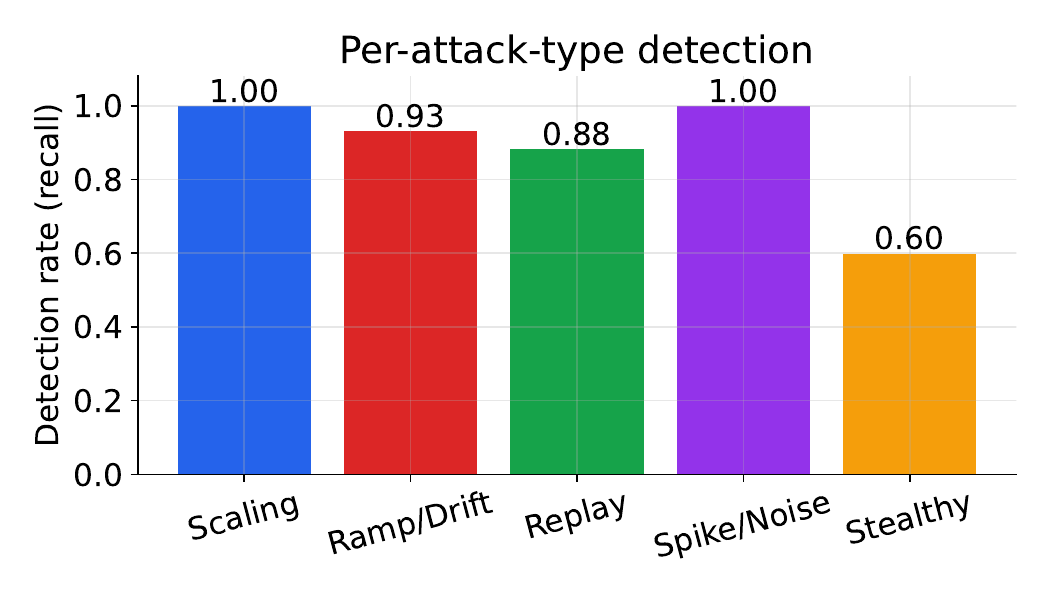}
    \caption{Per-attack-type detection rate (recall). Scaling and spike/noise are caught perfectly; stealthy FDIA achieves $60\%$ recall.}
    \vspace{-10pt}
    \label{fig:perattack}
    \vspace{-10pt}
\end{figure}

Per-attack-type recall is shown in Fig.~\ref{fig:perattack}. Scaling and spike/noise attacks are detected at $100\%$ recall, ramp/drift at $93\%$, and replay at $88\%$. The stealthy FDIA achieves $60\%$ recall, making it the most challenging attack type in the evaluated set. Nevertheless, the combined forecast-reconstruction score detects a meaningful fraction of small-bias attacks despite their close similarity to normal load behavior.

\medskip
\subsubsection{Privacy--Utility Trade-Off and Robustness}
\begin{figure}[!t]
    \centering
    \includegraphics[width=\columnwidth]{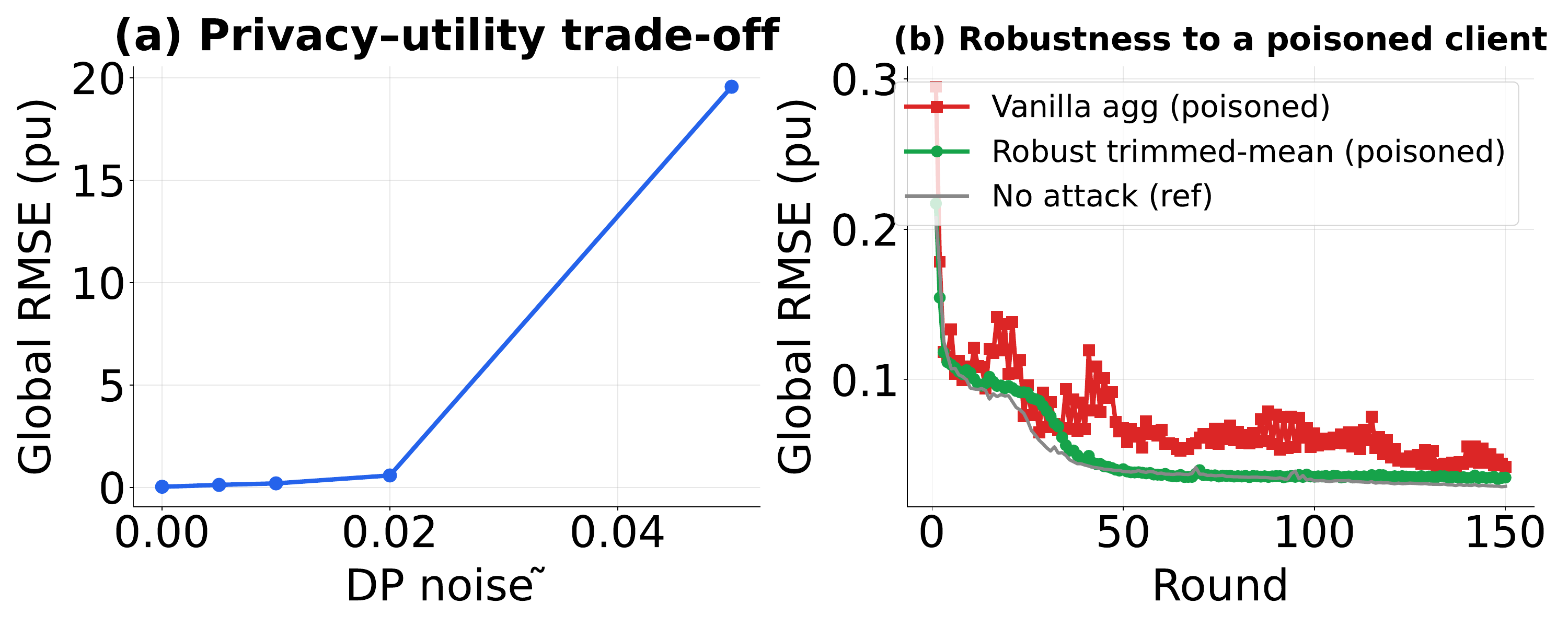}
    \caption{(a) Privacy--utility trade-off as the DP noise multiplier $\sigma$ increases from 0 to 0.05. (b) Robustness to a poisoned controller.}
    \vspace{-13pt}
    \label{fig:privrobust}
\end{figure}

Fig.~\ref{fig:privrobust}(a) shows the privacy--utility trade-off under different differential-privacy noise levels. Increasing the noise multiplier provides stronger update-level privacy but increases forecast error because useful information in the model updates is partially masked. Thus, the DP setting must be selected based on the required privacy--accuracy balance. Fig.~\ref{fig:privrobust}(b) shows robustness under a poisoned-client scenario. Vanilla aggregation is more affected by the malicious update, increasing the global RMSE to $0.042$~pu, while robust trimmed-mean aggregation reduces the impact to $0.035$~pu, closer to the no-attack reference.

\smallskip
These results show that \name can preserve local data privacy while improving resilience against compromised client updates. 
The proposed model achieves a 0.5~s-ahead RMSE of $0.023$--$0.038$~pu, compared with $0.32$--$0.34$~pu for persistence, indicating that the dual-head model can track rapid duty-cycling behavior more effectively than a last-value baseline. At the same time, the aggregate forecast provides operator-level visibility without exposing raw MW measurements from individual data centers.

The FDIA results confirm the benefit of combining forecasting and reconstruction errors. Obvious attacks such as scaling and spike/noise are detected reliably, while ramp/drift and replay attacks are also captured with high recall. Stealthy FDIA remains the most difficult case because its bias is close to normal load variation, but the calibrated score still provides useful local detection with high overall precision. The privacy and robustness studies further show the main deployment trade-off: differential privacy improves protection at the cost of accuracy, while robust trimmed-mean aggregation reduces the impact of a poisoned local update. 

\section{Conclusion}
\label{sec:conclusion}

This paper presented \namenospace, a federated framework for privacy-preserving load forecasting and local FDIA detection in data-center-rich power systems. The dual-head model combines forecasting and reconstruction while keeping raw measurements and absolute MW demand local. Validation using phasor-domain simulations of four data-center loads in the IEEE 39-bus system achieved a 0.5~s-ahead forecast RMSE of 0.023--0.038~pu and FDIA detection with ROC-AUC of 0.979, $F_1=0.930$, and precision of 0.988. Robust trimmed-mean aggregation further reduced the poisoned-client RMSE from 0.042 to 0.035~pu. The implementation is available at \url{https://github.com/KIBRIA-SAROARE/FedGuard-DC}. Future work will consider time-varying operating conditions, public cloud-workload traces, and \mbox{coordinated multi-DC attack studies.}


\section*{Acknowledgment}
This work was supported in part by the U.S. National Science Foundation under Grant OIA-2437963, the Louisiana Board of Regents, and the College of Engineering and Science at Louisiana Tech University.

\balance
\bibliographystyle{IEEEtran}
\bibliography{refs}

@techreport{nerc2025ltra,
  author       = {{North American Electric Reliability Corporation}},
  title        = {2025 Long-Term Reliability Assessment},
  institution  = {NERC},
  address      = {Atlanta, GA, USA},
  year         = {2025}
}

@inproceedings{mcmahan2017fedavg,
  author       = {McMahan, H. Brendan and Moore, Eider and Ramage, Daniel and Hampson, Seth and Ag\"{u}era y Arcas, Blaise},
  title        = {Communication-Efficient Learning of Deep Networks from Decentralized Data},
  booktitle    = {Proc. 20th Int. Conf. Artif. Intell. Statist. (AISTATS)},
  pages        = {1273--1282},
  year         = {2017}
}

@article{li2022secure,
  author       = {Li, Yang and Wei, Xinhao and Li, Yuanzheng and Dong, Zhaoyang and Shahidehpour, Mohammad},
  title        = {Detection of False Data Injection Attacks in Smart Grid: A Secure Federated Deep Learning Approach},
  journal      = {IEEE Trans. Smart Grid},
  volume       = {13},
  number       = {6},
  pages        = {4862--4872},
  month        = nov,
  year         = {2022}
}

@article{tran2023crosssilo,
  author       = {Tran, Hong-Yen and Hu, Jiankun and Yin, Xuechao and Pota, Hemanshu R.},
  title        = {An Efficient Privacy-Enhancing Cross-Silo Federated Learning and Applications for False Data Injection Attack Detection in Smart Grids},
  journal      = {IEEE Trans. Inf. Forensics Security},
  volume       = {18},
  pages        = {2538--2552},
  year         = {2023}
}

@article{lin2022incentive,
  author       = {Lin, Wen-Ting and Chen, Guo and Huang, Yi},
  title        = {Incentive Edge-Based Federated Learning for False Data Injection Attack Detection on Power Grid State Estimation: A Novel Mechanism Design Approach},
  journal      = {Appl. Energy},
  volume       = {314},
  pages        = {118828},
  month        = may,
  year         = {2022}
}

@article{lin2024privacy,
  author       = {Lin, Wen-Ting and Chen, Guo and Zhou, Xiaojun},
  title        = {Privacy-Preserving Federated Learning for Detecting False Data Injection Attacks on Power System},
  journal      = {Electr. Power Syst. Res.},
  volume       = {229},
  pages        = {110150},
  month        = apr,
  year         = {2024}
}

@article{kesici2024vertical,
  author       = {Kesici, Mert and Pal, Bikash and Yang, Guangya},
  title        = {Detection of False Data Injection Attacks in Distribution Networks: A Vertical Federated Learning Approach},
  journal      = {IEEE Trans. Smart Grid},
  volume       = {15},
  number       = {6},
  pages        = {5952--5964},
  month        = nov,
  year         = {2024}
}

@article{fernandez2022privacy,
  author       = {Fern{\'a}ndez, Joaqu{\'\i}n Delgado and Menci, Sergio Potenciano and Lee, Chul Min and Rieger, Alexander and Fridgen, Gilbert},
  title        = {Privacy-Preserving Federated Learning for Residential Short-Term Load Forecasting},
  journal      = {Appl. Energy},
  volume       = {326},
  pages        = {119915},
  month        = nov,
  year         = {2022}
}

@article{fekri2022rnn,
  author       = {Fekri, Mohammad Navid and Grolinger, Katarina and Mir, Saeed},
  title        = {Distributed Load Forecasting Using Smart Meter Data: Federated Learning with Recurrent Neural Networks},
  journal      = {Int. J. Electr. Power Energy Syst.},
  volume       = {137},
  pages        = {107669},
  month        = may,
  year         = {2022}
}

@article{athay1979ieee39,
  author       = {Athay, T. and Podmore, R. and Virmani, S.},
  title        = {A Practical Method for the Direct Analysis of Transient Stability},
  journal      = {IEEE Trans. Power App. Syst.},
  volume       = {PAS-98},
  number       = {2},
  pages        = {573--584},
  month        = mar,
  year         = {1979}
}

@article{dwork2014dp,
  author       = {Dwork, Cynthia and Roth, Aaron},
  title        = {The Algorithmic Foundations of Differential Privacy},
  journal      = {Found. Trends Theor. Comput. Sci.},
  volume       = {9},
  number       = {3--4},
  pages        = {211--407},
  year         = {2014}
}

@techreport{iea2025energyai,
  author      = {{International Energy Agency}},
  title       = {Energy and {AI}},
  institution = {IEA},
  address     = {Paris, France},
  year        = {2025},
  url         = {https://www.iea.org/reports/energy-and-ai}
}

@techreport{lbnl2024usdatacenter,
  author      = {Shehabi, Arman and Smith, Sarah J. and Hubbard, Ashley and Newkirk, Anna and Lei, Nian and Siddik, Md Abu Bakar and Holecek, Brian and Koomey, Jonathan and Masanet, Eric and Sartor, Dale},
  title       = {2024 United States Data Center Energy Usage Report},
  institution = {Lawrence Berkeley National Laboratory},
  address     = {Berkeley, CA, USA},
  number      = {LBNL-2001637},
  year        = {2024},
  url         = {https://eta.lbl.gov/publications/2024-lbnl-data-center-energy-usage-report}
}

@article{liu2011fdi,
  author  = {Liu, Yao and Ning, Peng and Reiter, Michael K.},
  title   = {False Data Injection Attacks against State Estimation in Electric Power Grids},
  journal = {ACM Trans. Inf. Syst. Secur.},
  volume  = {14},
  number  = {1},
  pages   = {13:1--13:33},
  year    = {2011},
  doi     = {10.1145/1952982.1952995}
}

@article{jithish2023distributed,
  author  = {Jithish, J. and Alangot, Bithin and Mahalingam, Nagarajan and Yeo, Kiat Seng},
  title   = {Distributed Anomaly Detection in Smart Grids: A Federated Learning-Based Approach},
  journal = {IEEE Access},
  volume  = {11},
  pages   = {7157--7179},
  year    = {2023},
  doi     = {10.1109/ACCESS.2023.3237554}
}

@article{li2024poisoning,
  author  = {Li, Xiumin and Wen, Mi and He, Siying and Lu, Rongxing and Wang, Liangliang},
  title   = {A Privacy-Preserving Federated Learning Scheme Against Poisoning Attacks in Smart Grid},
  journal = {IEEE Internet Things J.},
  volume  = {11},
  number  = {9},
  pages   = {16805--16816},
  year    = {2024},
  doi     = {10.1109/JIOT.2024.3365142}
}

@article{kececi2025localization,
  author  = {Ke{\c{c}}eci, Cihat and Davis, Katherine R. and Serpedin, Erchin},
  title   = {Federated Learning-Based Distributed Localization of False Data Injection Attacks on Smart Grids},
  journal = {IEEE Syst. J.},
  pages   = {1--11},
  year    = {2025},
  doi     = {10.1109/JSYST.2025.3581524}
}

@article{briggs2022residential,
  author  = {Briggs, Christopher and Fan, Zhong and Andras, Peter},
  title   = {Federated Learning for Short-Term Residential Load Forecasting},
  journal = {IEEE Open Access J. Power Energy},
  volume  = {9},
  pages   = {573--583},
  year    = {2022},
  doi     = {10.1109/OAJPE.2022.3206220}
}

@article{wang2023personalized,
  author  = {Wang, Yi and Gao, Ning and Hug, Gabriela},
  title   = {Personalized Federated Learning for Individual Consumer Load Forecasting},
  journal = {CSEE J. Power Energy Syst.},
  volume  = {9},
  number  = {1},
  pages   = {326--330},
  year    = {2023},
  doi     = {10.17775/CSEEJPES.2021.07350}
}

@article{qu2023heterogeneous,
  author  = {Qu, Xiaodong and Guan, Chengcheng and Xie, Gang and Tian, Zhiyi and Sood, Keshav and Sun, Chaoli and Cui, Lei},
  title   = {Personalized Federated Learning for Heterogeneous Residential Load Forecasting},
  journal = {Big Data Mining and Analytics},
  volume  = {6},
  number  = {4},
  pages   = {421--432},
  year    = {2023},
  doi     = {10.26599/BDMA.2022.9020043}
}

@article{husnoo2023secure,
  author  = {Husnoo, Muhammad Akbar and Anwar, Adnan and Hosseinzadeh, Nasser and Islam, Shama Naz and Mahmood, Abdun Naser and Doss, Robin},
  title   = {A Secure Federated Learning Framework for Residential Short-Term Load Forecasting},
  journal = {IEEE Trans. Smart Grid},
  volume  = {15},
  number  = {2},
  pages   = {2044--2055},
  month   = mar,
  year    = {2024},
  doi     = {10.1109/TSG.2023.3292382}
}

@inproceedings{saroare2026gridstream,
  author    = {Md Kibria Saroare and Md Abul Hasnat and Md Rubel Ahmed},
  title     = {{GridStream}: A Hardware-Efficient Framework for Bandwidth-Constrained Point-on-Wave Disturbance Monitoring},
  booktitle = {2026 IEEE Texas Power and Energy Conference (TPEC)},
  year      = {2026},
  doi       = {10.1109/TPEC67884.2026.11513170}
}

@misc{saroare2026dcclm,
  author        = {Saroare, Md Kibria and Ahmed, Md Rubel and Hussain, Arif},
  title         = {{DC-CLM}: Extending the {WECC} Composite Load Model for {AI} Data Center Dynamics},
  year          = {2026},
  eprint        = {2609.22093},
  archivePrefix = {arXiv},
  primaryClass  = {eess.SP},
  url           = {https://arxiv.org/abs/2609.22093}
}

\end{document}